\documentclass[fontsize=9pt]{llncs}
  \usepackage{microtype}
  \usepackage{amsmath,amsfonts}
  \usepackage[table]{xcolor}  \usepackage[T1]{fontenc}
  \usepackage[utf8]{inputenc}
  \usepackage{comment}
  \usepackage{forest}
  \usepackage{xspace}
  \usepackage{url}
  \usepackage{enumerate}
  \usepackage{listings}
  \usepackage{multirow}
  \usepackage{todonotes}
  \usepackage{thmtools}
  \usepackage{thm-restate}
  \usepackage[noend, boxed, noline]{algorithm2e}
  \usepackage[capitalise]{cleveref}
  \usepackage{graphics}
  \usepackage{tabularx}
  \usepackage{amssymb}
  \usepackage{epsfig}
  \usepackage[whileod]{chl-alg}
  \usepackage[caption=true,font=footnotesize]{subfig}
  \usepackage{verbatim}

  \usepackage{array}
\newcolumntype{C}[1]{>{\centering\let\newline\\\arraybackslash\hspace{0pt}}m{#1}}

\def\dd{\mathinner{.\,.}}
\newcommand{\cO}{\mathcal{O}}

 \newcommand{\defproblem}[3]{
  \vspace{2mm}
\noindent\fbox{
  \begin{minipage}{0.96\textwidth}
  #1\\
  {\bf{Input:}} #2  \\
  {\bf{Output:}} #3
  \end{minipage}
  }
  \vspace{2mm}
}

\begin{document}
\title{How to answer a small batch of RMQs or LCA queries in practice}

\author{
Mai Alzamel \inst{1}
\and
Panagiotis Charalampopoulos \inst{1}
\and
Costas~S.~Iliopoulos \inst{1}
\and
Solon~P.~Pissis \inst{1}
}

\institute{
    Department of Informatics, King's College London, UK\\
    \email{[mai.alzamel,panagiotis.charalampopoulos,\\costas.iliopoulos,solon.pissis]@kcl.ac.uk}\\[1ex]
   }

\date{}

\maketitle 
   
\begin{abstract}
In the Range Minimum Query (RMQ) problem, we are given an array $A$ of $n$ numbers and we are asked to answer queries of the following type: for indices $i$ and $j$ between $0$ and $n-1$, query $\text{RMQ}_A(i,j)$ returns the index of a minimum element in the subarray $A[i\dd j]$. Answering a small batch of RMQs is a core computational task in many real-world applications, in particular due to the connection with the Lowest Common Ancestor (LCA) problem. With {\em small batch}, we mean that the number $q$ of queries is $o(n)$ and we have them all at hand. It is therefore not relevant to build an $\Omega(n)$-sized data structure or spend $\Omega(n)$ time to build a more succinct one. It is well-known, among practitioners and elsewhere, that these data structures for online querying carry high constants in their pre-processing and querying time. We would thus like to answer this batch efficiently in practice. With {\em efficiently in practice}, we mean that we (ultimately) want to spend $n + \cO(q)$ time and $\cO(q)$ space. We write $n$ to stress that the number of operations per entry of $A$ should be a very small constant. Here we show how existing algorithms can be easily modified to satisfy these conditions. The presented experimental results highlight the practicality of this new scheme. The most significant improvement obtained is for answering a small batch of LCA queries.
A library implementation of the presented algorithms is made available.
\end{abstract}

\section{Introduction}

In the Range Minimum Query (RMQ) problem, we are given an array $A$ of $n$ numbers and we are asked to
answer queries of the following type: for indices $i$ and $j$ between $0$ and $n-1$, query $\text{RMQ}_A(i,j)$ returns the index of a minimum element in the subarray $A[i\dd j]$. 

The RMQ problem and the linearly equivalent Lowest Common Ancestor (LCA) problem~\cite{Bender2000} are very well-studied and several optimal algorithms exist to solve them. It was first shown by Harel and Tarjan~\cite{Tarjan84} that a tree can be pre-processed in $\cO(n)$ time so that LCA queries can be answered in $\cO(1)$ time per query. A major breakthrough in practicable constant-time LCA-computation was made by Berkman and Vishkin~\cite{Berkman93}. Farach and Bender~\cite{Bender2000} further simplified this algorithm by showing that the RMQ problem is linearly equivalent to the LCA problem (shown also in~\cite{Gabow84}). The constants due to the reduction, however, remained quite large, making these algorithms impractical in most realistic cases. To this end, Fischer and Heun~\cite{Fischer2006} presented yet another optimal, but also direct, algorithm for the RMQ problem. The same authors (but also others~\cite{DBLP:journals/jda/IlieNT10}) showed that due to large constants in the pre-processing and querying time implementations of this algorithm are often slower than implementations of the naive ones. Continuous efforts for engineering these solutions are being made~\cite{Ferrada2016}.

In this article we try to address this problem, in particular when one wants to answer a relatively small batch of RMQs efficiently. This version of the problem is a core computational task in many real-world applications such as in {\em object inheritance} during static compilation of code~\cite{Bender2005} or in several {\em string matching} problems (see Section~\ref{sec:apps} for some). With {\em small batch}, we mean that the number $q$ of the queries is $o(n)$ {\em and} we have them all at hand. It is therefore not relevant to build an $\Omega(n)$-sized data structure or spend $\Omega(n)$ time to build a more succinct one. It is well-known, among practitioners and elsewhere, that these data structures carry high constants in both their pre-processing and querying time. (Note that when $q=\Omega(n)$ one can use these data structures for this computation.) We would thus like to answer this batch efficiently in practice. With {\em efficiently in practice}, we mean that we (ultimately) want to spend $n + \cO(q)$ time and $\cO(q)$ space. We write $n$ to stress that the number of operations per entry of $A$ should be a very small constant; e.g.~scan the array once or twice. In what follows, we show how existing algorithms can be easily modified to satisfy these conditions. 
Experimental results presented here highlight the practicality of this scheme. The most significant improvement obtained is for answering a small batch of LCA queries. The RMQ Batch problem can be defined as follows.

\defproblem{RMQ Batch}{An array $A$ of size $n$ of numbers  and a list $Q$ of $q$ pairs of indices $(i,j)$, $0 \leq i \leq j \leq n-1$}{$\text{RMQ}_A(i,j)$ for each $(i,j) \in Q$}

\noindent The LCA Queries Batch problem can be defined as follows.

\defproblem{LCA Queries Batch}{A rooted tree $T$ with $n$ labelled nodes $0,1,\ldots,n-1$ and a list $Q$ of $q$ pairs of nodes $(u,v)$}{$\text{LCA}_T(u,v)$ for each $(u,v) \in Q$}

\noindent {\bf Our computational model.} We assume the word-RAM model with word size $w = \Omega(\log n)$. For the RMQ Batch problem, we assume that we are given a rewritable array $A$ of size $n$, each entry of which may be increased by $n$ and still fit in a computer word. 
For the LCA Queries Batch problem, we assume that we are given (an $\cO(n)$-sized representation of) a rewritable tree $T$ allowing constant-time access to (at least) the nodes of $T$ that are in some query in $Q$ (see the representation in~\cite{DBLP:journals/tcs/GearyRRR06}, for instance). All presented algorithms are deterministic. 

\section{Contracting the Input Array}\label{sec:con}

Consider any two adjacent array entries $A[i]$ and $A[i+1]$. Observe that if no query in $Q$ starts or ends at $i$ or at $i+1$ then, if $A[i] \neq A[i+1]$, $\max(A[i],A[i+1])$ will never be the answer to any of the queries in $Q$.
Hence, the idea is that we want to contract array $A$, so that each block that does not contain the left or right endpoint of any query gets replaced by one element: its minimum. A similar idea, based on sorting the list $Q$, has been considered in the {\em External Memory} model~\cite{DBLP:conf/swat/AfshaniS14} (see also~\cite{DBLP:conf/wads/ArgeFSS13}). In this section, we present a solution for our computational model, which avoids using $\Omega(n)$ space or time, but also avoids using $\Omega(\text{sort}(Q))$ time.

There are some technical details in order to update the queries for $A$ into queries for the new array using only $\cO(q)$ time and extra space.
We first scan the array $A$ once and find $\mu=\max_{i}A[i]$. 
We also create two auxiliary arrays $Z_0[0\dd 2q-1]$ and $Z_1[0\dd 2q-1]$.
For each query $(i,j) \in Q$ we mark positions $i$ (and $j$) in the array $A$ as follows. If $A[i] \leq \mu$, then $i$ has not been marked before. Let this be the $k$-th position, $k>0$, that gets marked (we just store a counter for that).
We store $A[i]$ in $Z_0[\mu+k \mod 2q]$ and replace the value that is stored in $A[i]$ by $\mu+k$. 
We also start a linked list at $Z_1[\mu+k \mod 2q]$, where we insert a pointer to query $(i,j)$, so that we can update it later. If $A[i] > \mu$, then the position has already been marked; we just add a pointer to the respective query in the linked list starting at $Z_1[A[i] \mod 2q]$.

We then scan array $A$ again and create a new array $A_Q$ as follows: for each marked position $j$ (i.e. $A[j]>\mu$), we copy the original value (i.e. $Z_0[A[j] \mod 2q]$) in $A_Q$, while each maximal block in $A$ that does not contain a marked position is replaced by a single entry---its minimum. 
When we insert the original entry of a marked position $j$ of $A$ (i.e. $Z_0[A[j] \mod 2q]$) in $A_Q$ at position $p$, we go through the linked list that is stored in $Z_1[A[j] \mod 2q]$, where we have stored pointers to all the queries of the form $(i,j)$ or $(j,k)$, and replace $j$ by $p$ in each of them.
Thus, after we have scanned $A$, for each query $(i,j) \in Q$ on $A$, we will have stored the respective pair $(i',j')$ on $A_Q$. Note that we need to scan array $A$ only {\em once} if we know $\mu$ a priori (e.g.~in LCP array~\cite{CHL07}), or {\em twice} otherwise.

\begin{example}
Assume we are given array $A$ and $Q=\{(4,18),(0,6),(6,10)\}$.
%
	\begin{center}
        $A~$\begin{tabular} {|C{0.4cm}|C{0.4cm}|C{0.4cm}|C{0.4cm}|C{0.4cm}|C{0.4cm}|C{0.4cm}|C{0.4cm}|C{0.4cm}|C{0.4cm}|C{0.4cm}|C{0.4cm}|C{0.4cm}|C{0.4cm}|C{0.4cm}|C{0.4cm}|C{0.4cm}|C{0.4cm}|C{0.4cm}|C{0.4cm}|C{0.4cm}|C{0.4cm}|}
\hline
{\cellcolor{gray}17} & 22 & 38 & 4 & {\cellcolor{gray}5} & 8 & {\cellcolor{gray}2} & 8 & 9 & 21 & {\cellcolor{gray}0} & 12 & 8 & 7 & 13 & 3 & 6 & 14 & {\cellcolor{gray}1} & 36 & 0 & 4 \\ \hline
\end{tabular}
\newline
\linebreak
	\end{center}
%
Then $A_Q$ is as follows.
%
	\begin{center}
    $A_Q~$
        \begin{tabular}{|C{0.4cm}|C{0.4cm}|C{0.4cm}|C{0.4cm}|C{0.4cm}|C{0.4cm}|C{0.4cm}|C{0.4cm}|C{0.4cm}|C{0.4cm}|C{0.4cm}|C{0.4cm}|}
\hline
${\cellcolor{gray}17}$ & $4$ & ${\cellcolor{gray}5}$ & $8$ & ${\cellcolor{gray}2}$ & $8$ & ${\cellcolor{gray}0}$ & $ 3 $ & ${\cellcolor{gray}1}$ & $0$ \\ \hline
\end{tabular}
\newline
\linebreak
	\end{center}
\end{example}

While creating $A_Q$, we also store in an auxiliary array the function $f:\{0,1,\ldots,|A_Q|-1\} \rightarrow \{0,1,\ldots,n-1\}$ between positions of $A_Q$ and the respective original positions in $A$. 

Now notice that $A_Q$ and the auxiliary arrays are all of size $\cO(q)$ since in the worst case we mark $2q$ distinct elements of $A$ and contract $2q+1$ blocks that do not contain a marked position. (We can actually throw away everything before the first marked position and everything after the last marked position and get $4q-1$ instead.) The whole procedure takes $n+\cO(q)$ time and $\cO(q)$ space. Note that if $\text{RMQ}_{A_Q}(i',j')=\ell$ then $\text{RMQ}_{A}(i,j)=f(\ell)$.

We can finally retrieve the original input array if required by replacing $A[f(j)]$ by $A_Q[j]$ for every $j$ in the domain of $f$ in $\cO(q)$ time.

\section{Small RMQ Batch}

\subsection{An $n + \cO(q \log q)$-time and $\cO(q)$-space Algorithm}

The algorithm presented in this section is a modification of the {\em Sparse Table} algorithm by Bender and Farach-Colton~\cite{Bender2000} applied on array $A_Q$; we denote it by $\textsf{ST-RMQ}$. The modification is based on the fact that {\bf (i)} we do not want to consume $\Omega(q \log q)$ extra space to answer the $q$ queries; and {\bf (ii)} we do not want to necessarily do all the pre-processing work of the algorithm in~\cite{Bender2000}, which is designed to answer any of the $\Theta(q^2)$ possible queries online.  We denote this modified algorithm by $\textsf{ST-RMQ}_{\textsf{CON}}$ and formalise it below.

\medskip

\begin{algo}{$\textsf{ST-RMQ}_{\textsf{CON}}$}{$$A$, $Q$$}
\SET{A_Q}{\Call{Contract}{A,Q}}
\ACT{\text{Store function $f$; store $(i',j')$ for every $(i,j)\in Q$}}
\DOFOR{\mbox{each $(i,j) \in Q$}}
	\IF{i=j}
		\CALL{Report}{(i,i),i}
	\ELSE
		\ACT{\mbox{Add $(i,j)$ in bucket $B_{\lfloor \log (j'-i') \rfloor}$}}
	\FI
\OD
\SET{t}{\max \{r | B_r \neq \emptyset \}+1}
\DOFOR{\mbox{$m=0$ to $|A_Q|-1$}}
	\SET{D[m]}{(A_Q[m],m)}
\OD
\DOFOR{\mbox{$k=0$ to $t-1$}}
	\DOFOR{\mbox{each $(i,j)\in B_k$}}
		\SET{(a,p)}{\min(D[i'],D[j'-2^k+1])}
		\CALL{Report}{(i,j), f(p)}
	\OD
	\DOFOR{\mbox{$m=0$ to $|A_Q|-1$}}
    \IF{m+2^k \leq |A_Q|-1}
		\SET{D[m]}{\min (D[m],D[m+2^k])}
    \FI
	\OD
\OD
\end{algo}

\medskip

The idea is to first put each $(i,j) \in Q$ with $i \neq j$ in a bucket $B_k$ based on the $k$ for which $2^k \leq j'-i' < 2^{k+1}$---we can have at most $\lceil \log (|A_Q|-1) \rceil$ such buckets. In this process, if we find queries of the form $(i,i) \in Q$, we answer them on the spot. We can do this in $\cO(q)$ time.

We then create an array $D$ of size $|A_Q|$ where we will store $2$-tuples $(a,p)$. 
In Step $k$, $D[m]$ will store the minimum value across $A_Q[m \dd m+2^k-1]$, as well as the position $p$, $m \leq p < m+2^k$ where it occurs. 
We initialise it as $D[m]=(A_Q[m],m)$ and we will then update it by utilising the {\em doubling technique}. At Step $0$ we answer all (trivial) queries that are stored in $B_0$; they are of the form $(i,i+1)$ and the answer can be found by looking at $\min(D[i'],D[i'+1])$---note that we compare elements of $D$ lexicographically. When we are done with $B_0$ we have to update $D$ by setting $D[m]=\min (D[m],D[m+2^0])$ for all $m<|A_Q|-1$.

Generally, in Step $k$, we answer the queries of $B_{k}$ as follows.
For query $(i,j)$, we find the answer by obtaining $\min(D[i'],D[j'-2^k+1]=(a,p)$. We then return $f(p)$. The point is that $\{i',\ldots,i'+2^k-1\} \cup \{j'-2^k+1,\ldots,j'\} = \{i',\ldots,j'\}$.
When we are done with $B_k$ we set $D[m]=\min (D[m],D[m+2^k])$ if $m+2^k \leq |A_Q|-1$. 

We do this until we have gone through all $t$ non-empty buckets (i.e.~$t=\max \{r | B_r \neq \emptyset \}+1$). Updating $D$ takes $\cO(q)$ time in each step, and we need in total $\cO(q)$ time for the queries. We thus need $\cO(q t)$ time for this part of the algorithm. Since $t=\max\{\lfloor \log (j'-i')\rfloor|(f(i'),f(j')) \in Q)\} = \cO(\log q)$, this time is $\cO(q \log q)$. 
The overall time complexity of the algorithm is thus $n + \cO(q \log q)$. Notably, the space required is only $\cO(q)$ as we overwrite $D$ in each step.

\subsection{$n + \cO(q)$-time and $\cO(q)$-space Algorithms}

{\em Offline-based algorithm.} Given an array $A$ of $n$ numbers its {\em Cartesian tree} is defined as follows. The root of the Cartesian tree is $A[i] = \min\{A[0],\ldots, A[n-1]\}$, its left subtree is computed recursively on $A[0],\ldots,A[i-1]$ and its right subtree on $A[i + 1],\ldots,A[n-1]$.
An LCA instance can be obtained from an RMQ instance on an array
$A$ by letting $T$ be the Cartesian tree of $A$ that can be constructed in $\cO(n)$ time~\cite{Gabow84}. It is easy to see that $\text{RMQ}_{A}(i, j)$ in $A$ translates to $\text{LCA}_{T}(A[i], A[j])$ in $T$.
The first step of this algorithm is to create array $A_Q$ in $n +\cO(q)$ time similarly to algorithm $\textsf{ST-RMQ}_{\textsf{CON}}$. The second step is to construct the Cartesian tree $T_Q$ of $A_Q$ in $\cO(q)$ time and extra space. Finally, we apply the offline algorithm by Gabow and Tarjan~\cite{DBLP:conf/stoc/GabowT83} to answer $q$ $\text{LCA}_{T_Q}$ queries in $\cO(q)$ time and extra space. This takes overall $n + \cO(q)$ time and $\cO(q)$ extra space. We denote this algorithm by $\textsf{OFF-RMQ}_{\textsf{CON}}$. We denote by $\textsf{OFF-RMQ}$ the same algorithm applied on array $A$.

{\em Online-based algorithm.} The first step of this algorithm is to create array $A_Q$ in $n +\cO(q)$ time similarly to algorithm $\textsf{ST-RMQ}_{\textsf{CON}}$. We can then apply the algorithm by Fischer and Heun~\cite{Fischer2006} on array $A_Q$ to obtain overall an $n + \cO(q)$-time and $\cO(q)$-space algorithm. We denote this algorithm by $\textsf{ON-RMQ}_{\textsf{CON}}$. We denote by $\textsf{ON-RMQ}$ the same algorithm applied on array $A$.

Note that in the case when $q = \Omega(n)$, i.e.~the batch is not so small, we can choose to apply algorithm $\textsf{OFF-RMQ}$ or algorithm $\textsf{ON-RMQ}$ on array $A$ directly thus obtaining an algorithm that always works in $n + \cO(q)$ time and $\cO(\min\{n,q\})$ extra space. We therefore obtain the following result asymptotically.

\begin{theorem}
The RMQ Batch problem can be solved in $n + \cO(q)$ time and $\cO(\min\{n,q\})$ extra space.
\end{theorem}

\section{Small LCA Queries Batch}

In the LCA problem, we are given a rooted tree $T$ having $n$ labelled nodes and we are asked to answer queries of the following type: for nodes $u$ and $v$, query $\text{LCA}_T(u,v)$ returns the node
furthest from the root that is an ancestor of both $u$ and $v$. There exists a time-optimal algorithm by Gabow and Tarjan~\cite{DBLP:conf/stoc/GabowT83} to answer a batch $Q$ of $q$ LCA queries in $\cO(n+q)$ time and $\cO(n)$ extra space. We denote this algorithm by $\textsf{OFF-LCA}$. In this section, we present a simple but {\em non-trivial} algorithm for improving this, for $q=o(n)$, to $n + \cO(q)$ time and $\cO(q)$ extra space. 

It is well-known (see~\cite{Bender2000} for the details) that an RMQ instance $A$ can be obtained from an LCA instance on a tree $T$ by writing down the depths of the nodes visited during an {\em Euler tour} of $T$. That is, $A$ is obtained by listing all node-visitations in a depth-first search (DFS) traversal of $T$ starting from the root. The LCA of two nodes translates to an RMQ (where we compare nodes based on their level) between the first occurrences of these nodes in $A$.

We proceed largely as in Section~\ref{sec:con}. For each query $(u,v) \in Q$, we mark nodes $u$ (and $v$) in $T$ as follows. If $u < n$ then $u$ has not been marked before. Let this be the $k$-th node, $k>0$, that gets marked (we just store a counter for that). We also create two arrays $Z_0[0\dd 2q-1]$ and $Z_1[0\dd 2q-1]$. We store $u$ in $Z_0[n - 1 + k \mod 2q]$ and replace $u$ by $n - 1 + k$. We also start a linked list at $Z_1[n - 1 + k \mod 2q]$, where we insert a pointer to query $(u,v)$, so that we can update it later. If $u > n-1$, the node has already been marked, and we just add a pointer to the respective query in the linked list starting at $Z_1[u \mod 2q]$.

We then do a single DFS traversal on $T$ and create two new arrays $E_Q$ and $L_Q$ as follows. When a marked node $v$ (i.e. $v > n - 1$) is visited for the {\em first time}, we write down in $E_Q$ its original value (i.e. $Z_0[v \mod 2q]$), while for each maximal sequence of visited nodes that are not marked we write down a single entry---the one with the {\em minimum tree level}. At the same time, we store in $L_Q[v]$ the level of the node added in $E_Q[v]$. While creating $E_Q$, we also store in an auxiliary array the function $f:\{0,1,\ldots,|E_Q|-1\} \rightarrow \{0,1,\ldots,n-1\}$ between positions of $E_Q$ and the respective node labels in $T$.

When we insert the original entry of a marked node $u$ of $T$ (i.e. $Z_0[u \mod 2q]$) in $E_Q$ at position $p$, we go through the linked list that is stored in $Z_1[u \mod 2q]$, where we have stored pointers to all the queries of the form $(u,v)$ or $(w,u)$, and replace $u$ by $p$ in each of these queries. Thus, after we have finished the traversal on $T$, for each LCA query $(u,v) \in Q$ on $T$, we will have stored the respective RMQ pair $(u',v')$ on $L_Q$; where $u'$ (resp.~$v'$) corresponds to the {\em first occurrence} of node $u$ (resp.~$v$) in the traversal. Thus we traverse $T$ only {\em once}.  

Now notice that $E_Q$ and the auxiliary arrays are all of size $\cO(q)$ since in the worst case we mark $2q$ distinct nodes of $T$ and contract $2q+1$ sequences of visited nodes that do not contain a marked node. (We can actually throw away everything before the first marked node and everything after the last marked node and get $4q-1$ instead.) The whole procedure takes $n+\cO(q)$ time and $\cO(q)$ space. We are now in a position to apply algorithm $\textsf{ON-RMQ}$ on $L_Q$ to obtain the final bound. To answer the queries, note that if $\text{RMQ}_{L_Q}(u',v')=\ell$ then $\text{LCA}_T(u,v)=E_Q[\ell]$.
We denote this algorithm by $\textsf{ON-LCA}_{\textsf{CON}}$. Alternatively, we can apply algorithm $\textsf{ST-RMQ}$ on $L_Q$ to solve this problem in $n+\cO(q \log q)$ and $\cO(q)$ extra space; we denote this algorithm by $\textsf{ST-LCA}_{\textsf{CON}}$. 

We can finally retrieve the original input tree if required by replacing node $f(v)$ by $E_Q[v]$ for every $v$ in the domain of $f$ in $\cO(q)$ time.

Note that in the case when $q = \Omega(n)$, i.e.~ the batch is not so small, we can choose to apply algorithm $\textsf{OFF-LCA}$ on tree $T$ directly, thus obtaining an algorithm that always works in $n + \cO(q)$ time and $\cO(\min\{n,q\})$ extra space. We therefore obtain the following result asymptotically.

\begin{theorem}
The LCA Queries Batch problem can be solved in $n + \cO(q)$ time and $\cO(\min\{n,q\})$ extra space.
\end{theorem}

\section{Applications}\label{sec:apps}

We consider the well-known application of answering $q$ LCA queries on the suffix tree of a string. 
The \textit{suffix tree} $T(S)$ of a non-empty string $S$ of length $n$ is a compact trie representing all suffixes of $S$ (see~\cite{CHL07}, for details). The nodes of the trie which become nodes of the suffix tree are called {\it explicit} nodes, while the other nodes are called {\it implicit}. 
Each edge of the suffix tree can be viewed as an upward maximal path of implicit nodes starting with an explicit node. Moreover, each node belongs to a unique path of that kind. Then, each node of the trie can be represented in the suffix tree by the edge it belongs to and an index within the corresponding path. The \textit{path-label} of a node $v$ is the concatenation of the edge labels along the path from the root to $v$. The nodes whose path-label corresponds to a suffix of $S$ are called {\em terminal}.
Given two terminal nodes $u$ and $v$ in $T(S)$, representing suffixes $S[i\dd n-1]$ and $S[j\dd n-1]$, the {\em string depth} of node $\text{LCA}_{T(S)}(u,v)$ corresponds to the {\em length} of their longest common prefix, also known as their longest common extension (LCE)~\cite{DBLP:journals/jda/IlieNT10}. 

In many textbook solutions for classical string matching problems (e.g.~maximal palindromic factors, approximate string matching with $k$-mismatches, approximate string matching with $k$-differences, online string search with the suffix array, etc.) we have that $q=\Omega(n)$ and/or the queries have to be answered {\em online}. In other algorithms, however, $q$ can be {\em much smaller} on average (in practice) and the queries can be answered {\em offline}. We describe here a few such solutions. The common idea, as in many fast average-case algorithms, is to minimise the number of queries by {\em filtering out} queries that can never lead to a valid solution.

\paragraph{Text indexing.}
Suppose we are given the suffix tree $T(S)$ of a text $S$ of length $n$ and we are asked to create the suffix links for the internal nodes. This may be necessary if the construction algorithm does not compute suffix links (e.g.~construction via suffix array) but they are needed for an application of interest. The \textit{suffix link} of a node $v$ with path-label $\alpha y$ is a pointer to the node path-labelled $y$, where  $\alpha \in \Sigma$ is a single letter and $y$ is a string. The suffix link of $v$ exists if $v$ is a non-root internal node of $T$. The suffix links can be computed as follows. The first step is to mark each internal node $v$ of the suffix tree with a pair of leaves $(i, j)$ such that leaves labelled $i$ and $j$ are in subtrees rooted at different children of $v$. This can be done by a DFS traversal of the tree. (Note that if an internal node $v$ has only one child then it must be terminal; assume that it represents the suffix $S[t \dd n-1]$. We thus create a suffix link to the node representing $S[t+1 \dd n-1]$.) Given an internal node $v$ marked with $(i, j)$, note that $v=\text{LCA}_{T(S)}(i, j)$, and let $\alpha y$ be its path-label. To create the suffix link from $v$, node $u$ with path-label $y$ can be obtained by the query $\text{LCA}_{T(S)}(i+1, j+1)$. We can create a batch of LCA queries consisting of all such pairs.
Note that in randomly generated texts, the number of internal nodes of $T(S)$ is $\cO(n/h)$ on average, where $h$ is the alphabet's entropy~\cite{DBLP:journals/tit/RegnierJ89}; thus the standard $\Theta(n)$-time and $\Theta(n)$-space solution to this problem, building the LCA data structure over $T(S)$~\cite{Bender2000}, is not satisfactory.

\paragraph{Finding frequent gapped factors in texts.}
We are given a text $S$ of length $n$, and positive integers $\ell_1$, $\ell_2$, $d$, and $k>1$.
The problem is to find all couples $(u,v)$, such that string $uwv$, for {\em any} string  $w$ (known as {\em gap} or {\em spacer}), $|w|=d$, occurs in $S$ at least $k$ times, $|u|=\ell_1$, $|v|=\ell_2$~\cite{iliopoulos:inria-00328042,MoTeX-II}. The first step is to build $T(S)$. We then locate all subtrees rooted at an explicit node with string depth at least $\ell_1$ and whose parent has string depth less than $\ell_1$, corresponding to factors $u$ repeated in $S$. From these subtrees, we only consider the ones with at least $k$ terminal nodes. Note that if $k$ is large enough, we may have only a few such subtrees. For each subtree with $k' \geq k$ terminal nodes, representing suffixes $S[i_{1}\dd n-1],S[i_{2}\dd n-1],\ldots,S[i_{k'}\dd n-1]$, we create a batch of LCA queries between all pairs $(i_{j}+\ell_1+d,i_{j'}+\ell_1+d)$
and report occurrences when LCA queries extend pairwise matches to length at least $\ell_2$ for a set of at least $k$ such suffixes. (This algorithm can be easily generalised for any number of gaps.) 

\paragraph{Pattern matching on weighted sequences.}
A {\em weighted sequence} specifies the probability of occurrence of each letter of the alphabet for every position. A weighted sequence thus represents many different strings,
each with the probability of occurrence equal to the product of probabilities of its letters
at subsequent positions of the weighted sequence.
The problem is to find all occurrences of a (standard) pattern $P$ of length $m$ with probability at least $1/z$ in a weighted sequence $S$ of length $n$~\cite{kociumaka_et_al:LIPIcs:2016:6816}. The first step is to construct the heavy string of $S$, denoted by $H(S)$, by assigning to $H(S)[i]$ the most probable letter of $S[i]$ (resolving ties arbitrarily). The second step is to build $T(P\$H(S))$, $\$ \notin \Sigma$. 
We can then compute the first mismatch between $P$ and every substring of $H(S)$. 
Note that the number of positions in $S$ where two or more letters occur with probability at least $1/z$ can be small, and so we consider only these positions to cause a legitimate mismatch between $P$ and a factor of $H(S)$. We then use $\cO(\log z)$ batches of LCA queries per such starting position to extend a match to length at least $m$. This is because $P$ cannot match a weighted sequence $S$ with probability $1/z$ if more than $\lfloor \log z \rfloor$ mismatches occur between $P$ and $H(S)$~\cite{kociumaka_et_al:LIPIcs:2016:6816}.

\paragraph{Pattern matching with don't care letters.}
We are given a pattern $P$ of length $m$, with $m-k$ letters from alphabet $\Sigma$ and 
$k$ occurrences of a don't care letter (matching itself and any letter from $\Sigma$), and a text $S$ of length $n$. The problem is to find all occurrences of $P$ in $S$~\cite{Pinter1985}. 
The first step is to build $T(P'\$S)$, $\$ \notin \Sigma$, where $P'$ is the string obtained from $P$ by replacing don't care letters with a letter $\# \notin \Sigma$. We then locate the subtree rooted at the highest explicit node corresponding to the longest factor $f$ of $P'$ without $\#$'s. We also locate, in the same subtree, all $V$ terminal nodes corresponding to starting positions of $f$ in $S$. Note that if $f$ is long enough, we may have only a few such nodes. Since we know where the don't care letters occur in $P$, we can create a batch of $kV$ LCA queries. An occurrence is then reported when LCA queries extend a match to length at least $m$. (This algorithm can be easily generalised for any number of patterns.)

\paragraph{Circular string matching.} We are given a pattern $P$ of length $m$ and a text $S$ of length $n$. The problem is to find all occurrences of $P$ or any of its cyclic shifts in $S$~\cite{circ}. The first step is to build $T(PP\$P^RP^R\#S\%S^{R})$, where $\$,\#,\% \notin \Sigma$, and $X^R$ denotes the reverse image of string $X$. We then conceptually split $P$ in two fragments of lengths $\lceil m/2 \rceil$ and $\lfloor m/2 \rfloor$. Any cyclic shift of $P$ contains as a factor at least one of the two fragments. We thus locate the two subtrees rooted at the highest explicit nodes corresponding to the fragments. We also locate in the same subtrees all $V$ terminal nodes corresponding to starting positions of the fragments in $S$. Note that if $m$ is long enough, we may have only a few such nodes. We create a batch of at most $2V$ LCA queries in order to extend to the left and to the right and report occurrences when LCA queries extend a match to length at least $m$. (This algorithm can be easily generalised for any number of patterns.)

\section{Experimental Results}

We have implemented algorithms $\textsf{ST-RMQ}_{\textsf{CON}}$, $\textsf{OFF-RMQ}_{\textsf{CON}}$, and $\textsf{ON-RMQ}_{\textsf{CON}}$ in the \texttt{C++} programming language. We have also implemented the same algorithms applied on the original array $A$, denoted by $\textsf{ST-RMQ}$, $\textsf{OFF-RMQ}$, and $\textsf{ON-RMQ}$, respectively; as well as the brute-force algorithm for answering RMQs in the two corresponding flavours, denoted by $\textsf{BF-RMQ}_{\textsf{CON}}$ and $\textsf{BF-RMQ}$. For the implementation of $\textsf{ON-RMQ}_{\textsf{CON}}$ and $\textsf{ON-RMQ}$ we used the \textsf{sdsl-lite} library~\cite{sdsl-lite}. If an algorithm requires $f(n,q)$ time and $g(n,q)$ extra space, we say that the algorithm has complexity $<f(n,q),g(n,q)>$. Table~\ref{tab:rmq-algos} summarises the implemented algorithms. The following experiments were conducted on a Desktop PC using one core of Intel Core i5-4690 CPU at 3.50GHz and 16GB of RAM. All programs were compiled with \texttt{g++} version 5.4.0 at optimisation level 3 (-O3).

\begin{figure}[!t]    
\centering
  \subfloat[$n=100,000,000$]
  {\label{fig:1a}\resizebox{.43\textwidth}{!}{\includegraphics[angle=270]{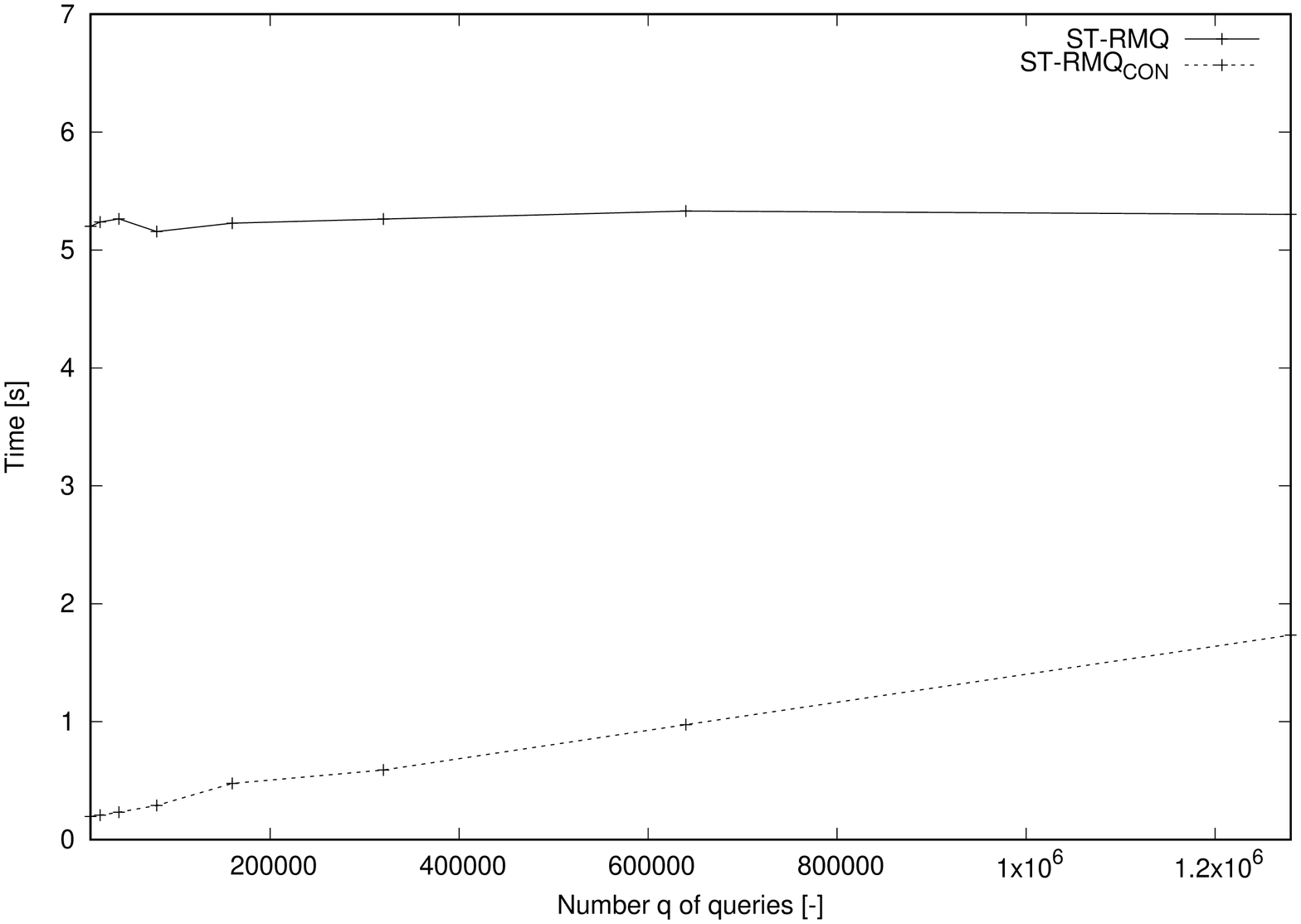}}}
  \subfloat[$n=100,000,000$]
  {\label{fig:1b}\resizebox{.43\textwidth}{!}{\includegraphics[angle=270]{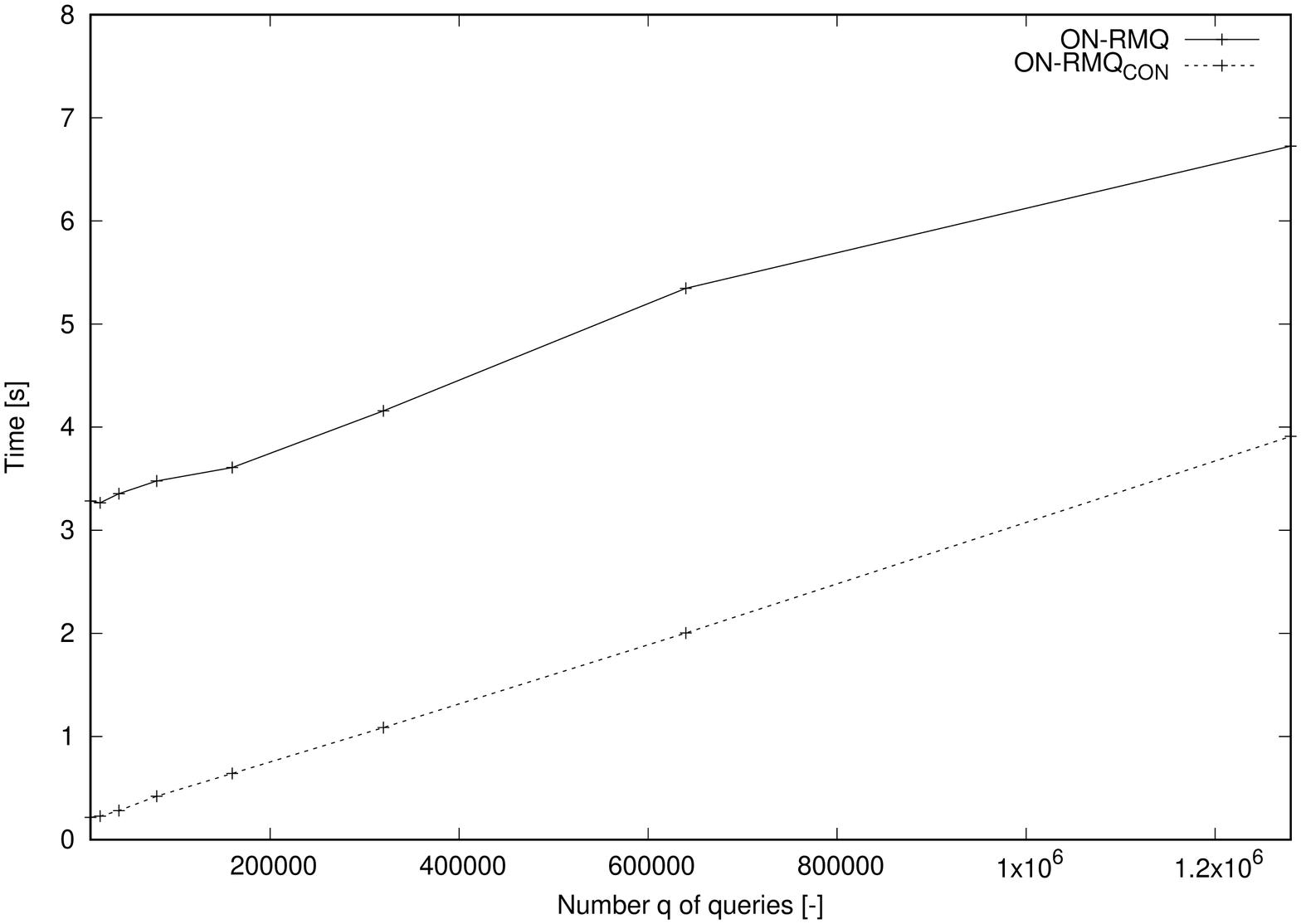}}}
  
  \subfloat[$n=1,000,000$]
  {\label{fig:1c}\resizebox{.43\textwidth}{!}{\includegraphics[angle=270]{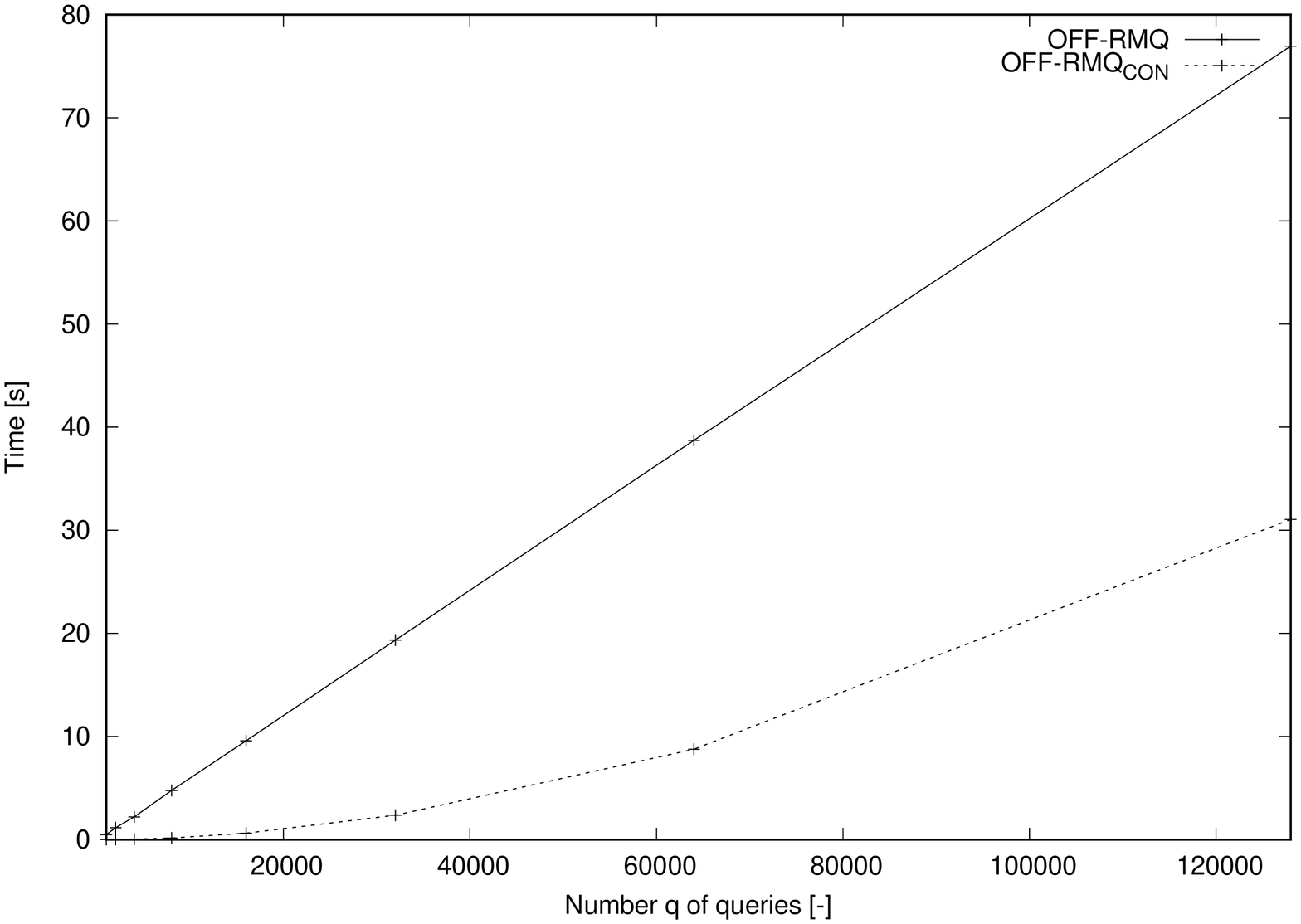}}}
  \subfloat[$n=1,000,000$]
  {\label{fig:1d}\resizebox{.43\textwidth}{!}{\includegraphics[angle=270]{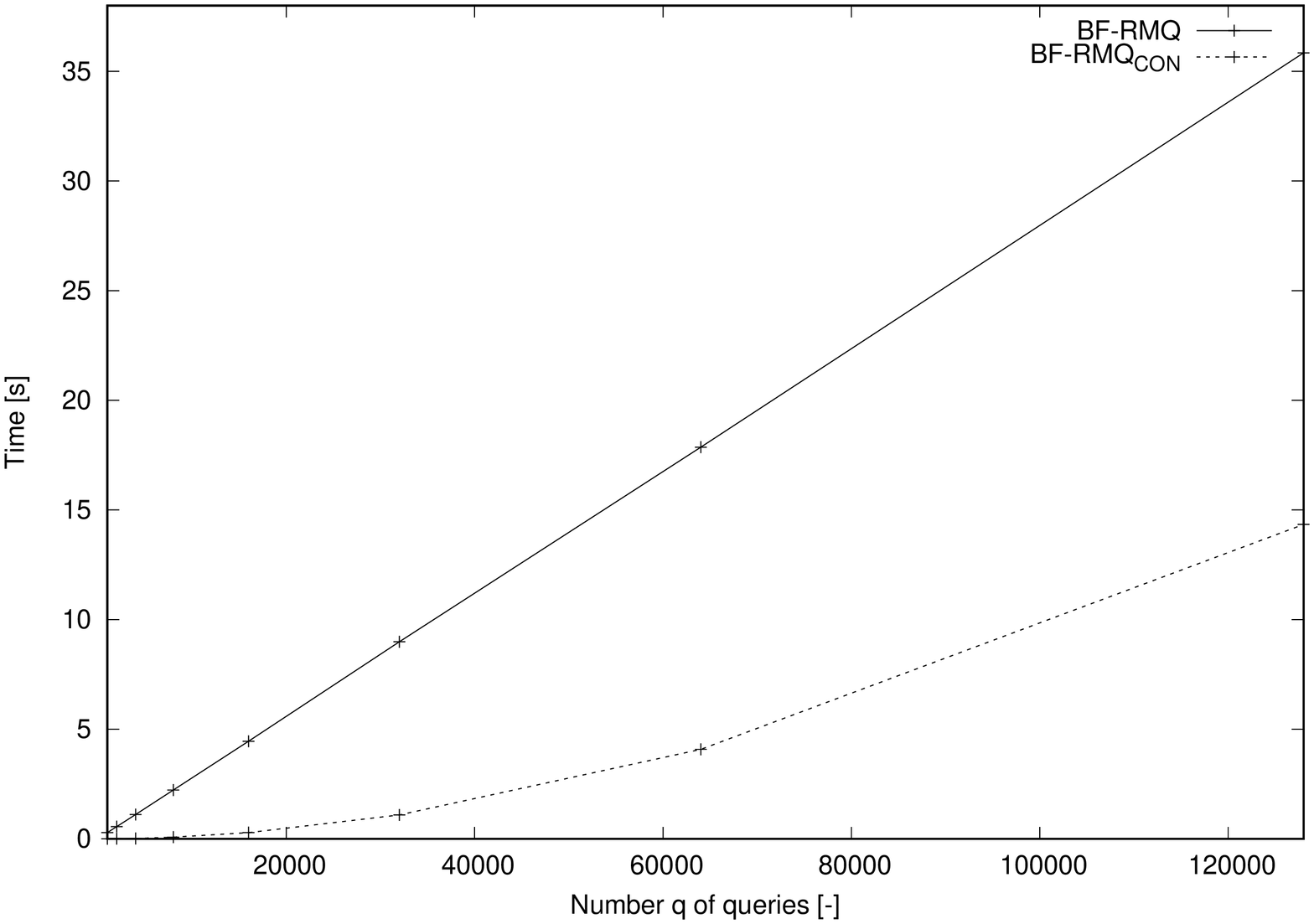}}}

    \caption{Impact of the proposed scheme on the RMQ algorithms of Table~\ref{tab:rmq-algos}.}
\label{fig:impact}
\end{figure}

\begin{table}[!t]
	\begin{center}
\begin{tabular} {|c|c|c|c|}
\hline

\multicolumn{2}{|c|}{\bf Non-Contracted}  &  \multicolumn{2}{|c|}{\bf Contracted}\\ \hline
 $\textsf{ST-RMQ}$ & $<\cO(n \log n+q),\cO(n \log n)>$ & $\textsf{ST-RMQ}_{\textsf{CON}}$ & $<n+\cO(q \log q),\cO(q)>$ \\ \hline
 $\textsf{ON-RMQ}$ & $<\cO(n+q),\cO(n)>$ &$\textsf{ON-RMQ}_{\textsf{CON}}$& $<n+\cO(q),\cO(q)>$\\ \hline
$\textsf{OFF-RMQ}$ & $<\cO(n+q),\cO(n)>$ &$\textsf{OFF-RMQ}_{\textsf{CON}}$& $<n+\cO(q),\cO(q)>$\\ \hline
$\textsf{BF-RMQ}$ & $<\cO(qn),\cO(1)>$ &$\textsf{BF-RMQ}_{\textsf{CON}}$ & $<n+\cO(q^2),\cO(q)>$\\ 
\hline
\end{tabular}
\newline
\linebreak
   \caption{Time and space complexities of algorithms for answering RMQs offline.}
	\label{tab:rmq-algos}
	\end{center}
\end{table}
{\em Experiment I.} We generated random (uniform distribution) input arrays of $n=1,000,000$ and $n=100,000,000$ entries (integers), and random (uniform distribution) lists of queries of sizes varying from $\sqrt{n}$ to $128 \sqrt{n}$, doubling each time. We compared the runtime of the implementations of the algorithms in Table~\ref{tab:rmq-algos} on these inputs; in particular, for each algorithm, we compared the standard implementation against the one with the contracted array. 
We used the large array, $n=100,000,000$, for $\textsf{ST-RMQ}$ and $\textsf{ON-RMQ}$ because they are significantly faster and the small one, $n=1,000,000$, for $\textsf{OFF-RMQ}$ and $\textsf{BF-RMQ}$.
The results plotted in Figure~\ref{fig:impact} show that the proposed scheme of contracting the input array improves the performance for all implementations substantially. 

\begin{figure}[!t]    
\centering
  \resizebox{.70\textwidth}{!}{\includegraphics[angle=270]{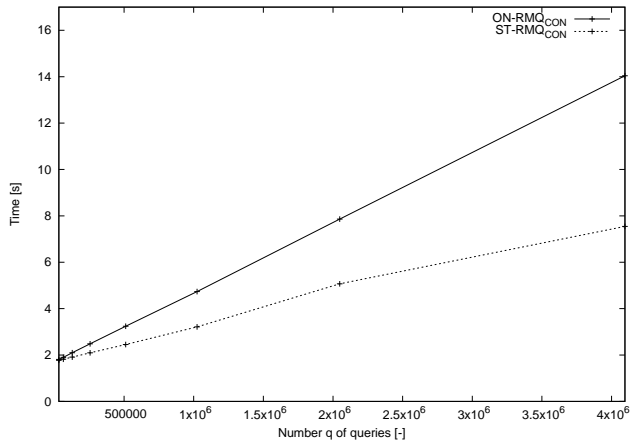}}
  \caption{Elapsed-time comparison of $\textsf{ON-RMQ}_{\textsf{CON}}$ and $\textsf{ST-RMQ}_{\textsf{CON}}$ algorithms for $n=1,000,000,000$.}
\label{fig:RMQcomparison}
\end{figure}

{\em Experiment II.} We generated random input arrays of $n=1,000,000,000$ entries, and random lists of queries of sizes varying from $\sqrt{n}$ to $128 \sqrt{n}$, doubling each time. We then compared the runtime of $\textsf{ON-RMQ}_{\textsf{CON}}$ and $\textsf{ST-RMQ}_{\textsf{CON}}$ on these inputs. The results are plotted in Figure~\ref{fig:RMQcomparison}. We observe that $\textsf{ST-RMQ}_{\textsf{CON}}$ becomes two times faster than $\textsf{ON-RMQ}_{\textsf{CON}}$ as $q$ grows. Notably, it was not possible to run this experiment with $\textsf{ON-RMQ}$, which implements a {\em succinct} data structure for answering RMQs, due to insufficient amount of main memory.

\begin{figure}[!t]    
\centering
  \resizebox{.70\textwidth}{!}{\includegraphics[angle=270]{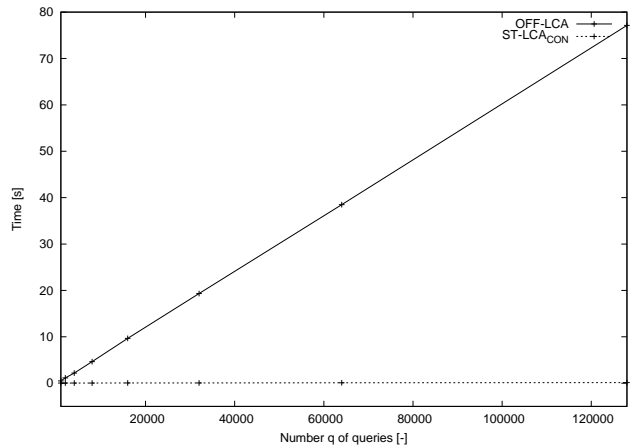}}
  \caption{Elapsed-time comparison of $\textsf{OFF-LCA}$ and $\textsf{ST-LCA}_{\textsf{CON}}$ algorithms for $n=1,000,000$.}
\label{fig:LCAcomparison}
\end{figure}

{\em Experiment III.} In addition, we have implemented algorithms $\textsf{ST-LCA}_{\textsf{CON}}$ and $\textsf{OFF-LCA}$ for answering LCA queries.  We first generated a random input array of $n=1,000,000$ entries and used this array to compute its Cartesian tree.
Next we generated random lists of LCA queries of sizes varying from $\sqrt{n}$ to $128 \sqrt{n}$, doubling each time. We then compared the runtime of $\textsf{OFF-LCA}$ and $\textsf{ST-LCA}_{\textsf{CON}}$ on these inputs. The results plotted in Figure~\ref{fig:LCAcomparison} show that the implementation of $\textsf{ST-LCA}_{\textsf{CON}}$ is more than two orders of magnitude faster than the implementation of $\textsf{OFF-LCA}$, highlighting the impact of the proposed scheme on LCA queries. 

\section{Final Remarks}

In this article, we presented a new family of algorithms for answering a small batch of RMQs or LCA queries in practice. The main purpose was to show that if the number $q$ of queries is small with respect to $n$ and we have them all at hand existing algorithms for RMQs and LCA queries can be easily modified to perform in $n + \cO(q)$ time and $\cO(q)$ extra space. The presented experimental results indeed show that with this new scheme significant practical improvements can be obtained; in particular, for answering a small batch of LCA queries. 

Specifically, algorithms $\textsf{ST-RMQ}_{\textsf{CON}}$ and $\textsf{ST-LCA}_{\textsf{CON}}$, our modifications to the {\em Sparse Table} algorithm whose main catch is $\Theta(n \log n)$ space~\cite{Bender2000}, seem to be the best way to answer in practice a small batch of RMQs and LCA queries, respectively. A library implementation of $\textsf{ST-RMQ}_{\textsf{CON}}$ is available at \url{https://github.com/solonas13/rmqo} under the GNU General Public License.  

\bibliographystyle{abbrv}
\bibliography{references}

\end{document}